# Surface properties of semiconductors from post-illumination photovoltage transient

Yury Turkulets,[1] and Ilan Shalish[1]*
[1]School of Electrical Engineering, Ben-Gurion University, Beer Sheva 8410501, Israel.

Free surfaces of semiconductors respond to light by varying their surface voltage (surface band bending). This surface photovoltage may be easily detected using a Kelvin probe. Modeling the transient temporal behavior of the surface photovoltage after the light is turned off may serve as a means to characterize several key electronic properties of the semiconductor, which are of fundamental importance in numerous electronic device applications, such as transistors and solar cells. In this paper, we develop a model for this temporal behavior and use it to characterize layers and nanowires of several semiconductors. Our results suggest that what has previously been considered to be a logarithmic decay is only approximately so. Due to the known limited frequency bandwidth of the Kelvin probe method, most previous Kelvin-probe-based methods have been limited to "slow responding" semiconductors. The model we propose extends this range of applicability.

## I. INTRODUCTION

Semiconductor free surfaces respond to light in the formation of surface photovoltage. The common scenario in most semiconductor surfaces is the trapping of majority carriers in surface states which gives rise to band-bending at the surface. The main effect of illumination is an internal photo-emission of these trapped charges over the surface barrier into the bulk. In this process, the density of the surface charge is reduced, reducing the band-bending, and this change in the surface band-bending constitutes the photovoltage.[1] In the following discussion, we will assume that the photon energy used in this process is not sufficient to cause band-to-band transitions, i.e., smaller than the forbidden energy gap. This way, we avoid a contribution of other mechanisms that may contribute photovoltage, such as the Dember effect.[2] Later on, we will show that this assumption may be removed for the method we propose. We will also assume a single type and single distribution of surface state. We will discuss later why an additional surface state may be ignored in many cases when using the proposed method to measure the equilibrium surface band bending. In the following, we will also limit the discussion to the free surface of n-type semiconductors, although the same should be generally applicable to any semiconductor-insulator junction of both conductivity types.

A semiconductor free surface may be viewed as a charged capacitor, where charges trapped in surface states, forming one side of the capacitor, are balanced by an adjacent surface depletion region (Fig. 1a). Direct current through a capacitor cannot be sustained for long. However, one may obtain a transient current by discharging the capacitor, or, in the case discussed here, by discharging surface traps, and monitoring the transient response as majority carriers are being re-trapped (Fig.1b). To measure the true properties of a free surface, one needs to avoid any direct contact with the surface, i.e., the surface must remain absolutely free. An easy way to discharge surface states is by photon absorption to excite electrons from surface traps into the conduction band. Monitoring the current *directly* is not possible without a metal contact, but it may be monitored *indirectly* by monitoring the corresponding change in the surface band banding using a Kelvin probe. This way, the built-in voltage is reduced by the *photovoltage, $V_{PV}$*, rather than by an externally applied voltage. After the light is turned off, the photovoltage gradually subsides over time to zero, as charge carriers return to the surface traps. Photovoltage may be conveniently monitored during this process using a Kelvin probe.[1]

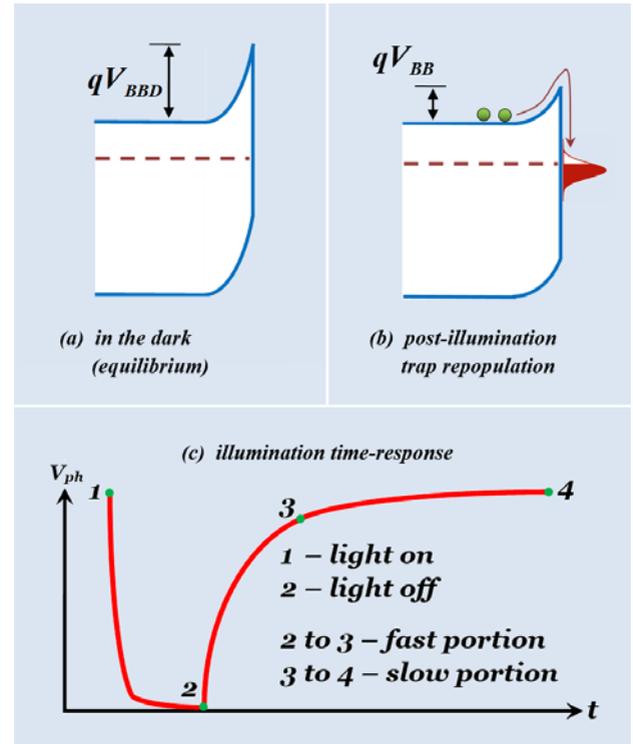

FIG 1 Band diagram of surface depleted n-type semiconductor (a) In the dark, before irradiation. A Gaussian drawn on the surface represents an energy distribution of surface states that occupy trapped electrons. The trapped electrons repel mobile electron creating an electric field and bending the bands. (b) After irradiation with photons of energy greater than the bandgap, the trap is partly depopulated, the electric field is smaller, and so is the band bending. Electrons are shown to traverse the built-in barrier from the conduction band to the surface to repopulate the traps. (c) The photovoltage time-response.

Since the discharge is carried out by energetic photons, it is typically much faster than the inverse process of repopulation which takes place in the dark using an orders-of-magnitude-smaller phonon energy. Furthermore, the typically slow repopulation process may be described as comprised of a short "fast" phase followed by a long slower phase (Fig. 2).[3] Several attempts have been made so far to model the decay of the photovoltage after turning off the light.[4,5,6,7,8,9,10,11,12,13,14] Two of these also propose methods based on their models to calculate the equilibrium band bending using derivatives of the light-on/ light-off responses.[10,14] These derivatives are obtained from the "fast" parts of the responses. While these methods appear to be correct and valid, their applicability using a Kelvin probe has been generally criticized as being limited to materials, which response is slow enough to fit within the frequency bandwidth of the Kelvin probe.[1] As we will show next, to avoid such use of derivatives would require to solve a non-analytic non-linear differential equation. Most of the authors do reach the same equation, but omit certain critical parts of it in order to solve it analytically, and commonly reach an approximate solution in the form of a *logarithmic decay*.

In this paper, we present a method to evaluate the equilibrium band bending without using derivatives, in a way that is not limited by the Kelvin probe band width. To this end, we solve the non-analytic differential equation describing the photovoltage decay, both for bulk layers and for nanowires, and use it as a model to extract the equilibrium band-bending, and consequently, several other related electrical properties of the semiconductor free surface.

II. MODEL

Since most of the previous models quite unanimously reach the same equation, we could actually start from that equation. However, for the completeness of this manuscript, we describe our way to reach the same equation. While it is possible to start from writing rate equations for the process, we realized that this has already been carried out and used successfully for Schottky barriers by Bethe. [15] Here, we start from the thermionic emission model of Bethe and introduce the necessary modifications.

Schottky barrier was originally defined to describe the unipolar electrostatic barrier formed at the metal-semiconductor junction,[16] but has been extended later to include the same type of barrier found in semiconductor heterojunctions[17,18] and even certain homojunctions, e.g. grain boundary junction,[19,20] all of them conductive junctions. As a matter of fact, the same type of barrier is present in semiconductor-insulator junctions as well.[21] One major difference from the Schottky barrier is that the free surface barrier does not remain altogether constant in our process. Therefore, we will have to find out the range, over which the assumption of constant barrier will be valid. This same validity verification may actually be required for all the other existing methods as well, because regardless of the way they were reached, they practically all reach the same equation.

During the surface state repopulation process, charges flow back from the semiconductor into the surface traps. The direction of the current is equivalent to that in a forward-biased Schottky diode. Most of this flow process is assumed to take place by means of thermionic emission over the barrier. As we discuss later, this assumption may not be valid for very small built-in fields. Using Bethe's thermionic emission model,[22] the (forward) current density may be described by

$$J = C_1 \cdot P_e \cdot exp\left(-\frac{\phi_B}{kT}\right) exp\left(\frac{qV_A}{kT}\right)=$$
$$= C_1 \cdot P_e \cdot exp\left(-q\frac{V_n + V_{BB}}{kT}\right) exp\left(\frac{qV_{PV}}{kT}\right) \qquad (1)$$

where $\phi_B$ is the Schottky barrier height, $V_A$ is the applied voltage, and $C_1 = 4\pi qm(kT)^2/h^3$, where q – electron charge, m – electron effective mass, h – Planck constant, k – Boltzmann constant, T – absolute temperature. In a free surface, the equivalent of the Schottky barrier $\phi_B$ is the surface barrier, which is comprised of $V_{BB}$ – the equilibrium band bending, and $V_n$ – the difference between the Fermi level and the conduction band minimum in the bulk. In our experiment, the photovoltage, $V_{PV}$, replaces the externally applied voltage, $V_A$. On polar faces of polar semiconductors, $V_{BB}$ is composed of two components. One is the band bending induced by charged surface state traps, and another, unique to polar materials, is a band bending induced by polar charge on the polar faces. While surface traps may be optically discharged, polar charges are constant and are not affected by light.

We also introduce the parameter $P_e$ for the electron emission probability. Essentially, this is the probability that an electron can find an unoccupied electronic state to transfer into on the other side of the barrier. In the case of a metal-semiconductor contact, the metal covers the entire surface area and therefore, and it is commonly assumed that electronic states in the metal are available for the thermally emitted electron anywhere on the surface, which renders this probability equal to 1. However, in the case of the free surface, where an electron is emitted into a surface state, this probability is smaller than 1, because surface states are not uniformly distributed, and because each surface state is only associated with a limited area, within which a passing electron can be trapped – the electron capture cross-section, $\sigma_n$. Therefore, if the density of charged surface states in equilibrium is $N_{TD}$, while the density of unoccupied state in equilibrium is $(B - 1) \cdot N_{TD}$, and if $N_T(t)$ is the density of charged surface states at a time $t$ after turning the light off, then the capture probability is the product of the electron capture cross-section, $\sigma_n$, and the density of unoccupied states available for trapping

$$P_e = \sigma_n[B \cdot N_{TD} - N_T(t)] \qquad (2)$$

The availability of empty states clearly varies with the photovoltage along the process.

Since the current density equals the change in the surface charge density, $qN_T(t)$, we can also write:

$$J = \frac{1}{2}q\frac{dN_T(t)}{dt} \qquad (3)$$

The factor of ½ is because the space charge region is not of constant capacitance but varies along the process as well. We note that the surface charge may include, in the case of polar materials, a time-invariant component of the polar charge.

The built-in field in the depletion region may be obtained from a solution of Poisson's equation. The maximum of the built-in electric field is reached at the very surface and is given by:[23]

$$E(t) = \sqrt{\frac{2qN_D}{\varepsilon}(V_{BB} - V_{PV})} = \frac{q}{\varepsilon}N_T(t) \qquad (3)$$

We note that $N_T(t)$ may include a time-invariant component, $N_{Polar}$, which is the polar charge on polar faces of polar materials.[24] For non-polar materials (or non-polar faces of polar materials), $N_{Polar}=0$. For convenience, we define a dimensionless variable:

$$x = 1 - \frac{V_{PV}}{V_{BB}} \qquad (4)$$

Extracting the surface charge density, $N_T$, from Eq. 3, we get

$$N_T(t) = \sqrt{\frac{2\varepsilon N_D}{q}(V_{BB} - V_{PV})} = N_{TD}\sqrt{x} \qquad (5)$$

Since the photovoltage may vary between $V_{BB}$ and zero, the variable $x$ may vary between 0 and 1. Taking the temporal derivative of Eq. 4, we get another expression for the current

$$J = \frac{q}{2}\frac{dN_T(t)}{dt} = \frac{q}{2}\frac{dN_T(t)}{dx}\frac{dx}{dt} = = \frac{qN_{TD}}{2\sqrt{x}}\frac{dx}{dt} \qquad (6)$$

which is equal to the current in Eq. 1. Expressing Eqs. 1 and 2 in terms of x, and equating Eqs. 1 and 6, we get

$$\frac{exp(Ax)}{(B - \sqrt{x})\sqrt{x}}\frac{dx}{dt} = \frac{1}{\tau} \qquad (7)$$

where

$$\tau = \frac{h^3 exp(qV_n/kT)}{8\pi\sigma_n m(kT)^2} = \frac{h^3 n_i exp(qE_g/2kT)}{8\pi\sigma_n N_D m(kT)^2}$$

$$and \quad A = \frac{qV_{BB}}{kT} \qquad (8)$$

The common practice in previous works has been to assume that the product $(B - \sqrt{x})\sqrt{x}$ is constant. Under such assumption a logarithmic decay is readily obtained. Equation 7 may be solved by separation of variables (a detailed solution is given in the Appendix), and the solution may be fitted to the photovoltage time-response data to yield the equilibrium surface band-bending, $V_{BB}$, the time constant $\tau$, and the parameter B. From $V_{BB}$, one can calculate the density of charged surface states, $N_{TD}$, and the total density of surface states, $B \cdot N_{TD}$. Given the doping concentration, $N_D$, the time constant may yield the capture cross-section, $\sigma_n$.

As noted above, Bethe's thermionic emission model requires a certain minimal barrier height. Immediately after the light is turned off (in the range noted as the "fast" range in Fig. 2), the barrier height is the lowest and may not always be enough to facilitate thermionic emission. This range is therefore not described by our model.

As surface photovoltage is contactless, it may conveniently be used on nanowires (or other nanostructure, or even powders) as well, without the need to fabricate metal contacts to individual wires. The measurement in this case integrates over an area containing a large number of wires. The above derivation is suitable for layer geometry. To use it on nano-structures, the specific structure has to be considered. For example, to characterize nanowires, we need to consider a cylindrical structure of radius R. [25] It can be shown that the electric field at the surface is given by

$$E(t) = \frac{2}{R}(V_{BB} - V_{PV}) = \frac{q}{\varepsilon}N_T(t) \qquad (10)$$

From Eq. 10, we get the surface charge

$$N_T(t) = \frac{2\varepsilon}{qR}(V_{BB} - V_{PV}) = N_{TD}x \qquad (11)$$

Changing variable to x and taking the time derivative of Eq. 11, we get an expression for the current density

$$J = \frac{q}{2}\frac{dN_T(t)}{dt} = \frac{q}{2}\frac{dN_T(t)}{dx}\frac{dx}{dt} = \frac{q}{2}N_{TD}\frac{dx}{dt} \qquad (12)$$

Since this current is equal to the current in Eq. 1, we get

$$\frac{dx}{dt} = \frac{1}{\tau_2}(B - \sqrt{x})exp(-x) \qquad (13)$$

Where

$$\tau_2 = \frac{h^3 n_i exp(E_g/2kT)}{8\pi\sigma_n m(kT)^2 N_D} \qquad (14)$$

Equation 13 may as well be solved by separation of variables (solution is detailed in the Appendix).

Since surface states are discharged when exposed to light, one should expect a certain modification of the height of the surface potential barrier. This is an aspect, in which the free surface is clearly different from the Schottky barrier treated by Bethe's thermionic emission model, for which the barrier may safely be assumed to be constant. However, as the experimental results show, most of this change takes place at the very beginning of the relaxation process, immediately after the light is turned off, while during the remaining part of the response (i.e., over the "slow" range), the surface barrier change appears to be negligible.

The solutions of Eqs. 7 and 13 may be used to fit the temporal response data of bulk layers or nanowires, respectively, to obtain the following parameters: (1) the equilibrium surface band-bending, $V_{BB}$, and the equilibrium density of surface charge, $N_{TD}$, (2) the density of surface states, $B \cdot N_{TD}$, (3) the time constant, $\tau$, and if $n_i$ and $N_D$ are known, the capture cross-section of surface states, $\sigma_n$.

III. EXPERIMENTAL DETAILS

(1) Materials

We tested the method on two materials. Our first choice was GaAs – a well-studied material. The n-GaAs wafer was obtained from AXT Inc. and was $2 \cdot 10^{17}\ cm^{-3}$ doped with S. The wafer was as-polished, cleaned sequentially with acetone and methanol and blown dry with nitrogen, immediately prior to measurement. We also wanted to present a case of a polar face of a polar semiconductor. To this end, we used an unintentionally doped ($5 \cdot 10^{16}\ cm^{-3}$) CdS (Eagle-Picher) with n-type conductivity. The same solvent cleaning procedure was used on this sample as well. To test the method on nanowires, we used hydride vapor phase epitaxy grown GaN nanowires on sapphire. Details of the growth and microscope images thereof may be found elsewhere.[26]

(2) Methods

Illumination was carried out using a ~50 nW of light from a monochromatized and filtered 300 W Xe lamp. To excite surface trapped charges, we used sub-bandgap wavelength illumination. However, as our model describes only the "slow" part of the photovoltage decay, there is practically no difference between below- or above-bandgap excitation. When using above bandgap excitation, electron-hole pairs will be excited along with surface trapped electrons, but the former will recombine and exit the scene shortly after the light is turned off, while the time constants for re-trapping will typically be several orders of magnitude slower. Moreover, there is no need to completely evacuate the surface state, and this means that the excitation may be short and partial. It also means that the photon energy does not need to cover the entire distribution of the surface state.

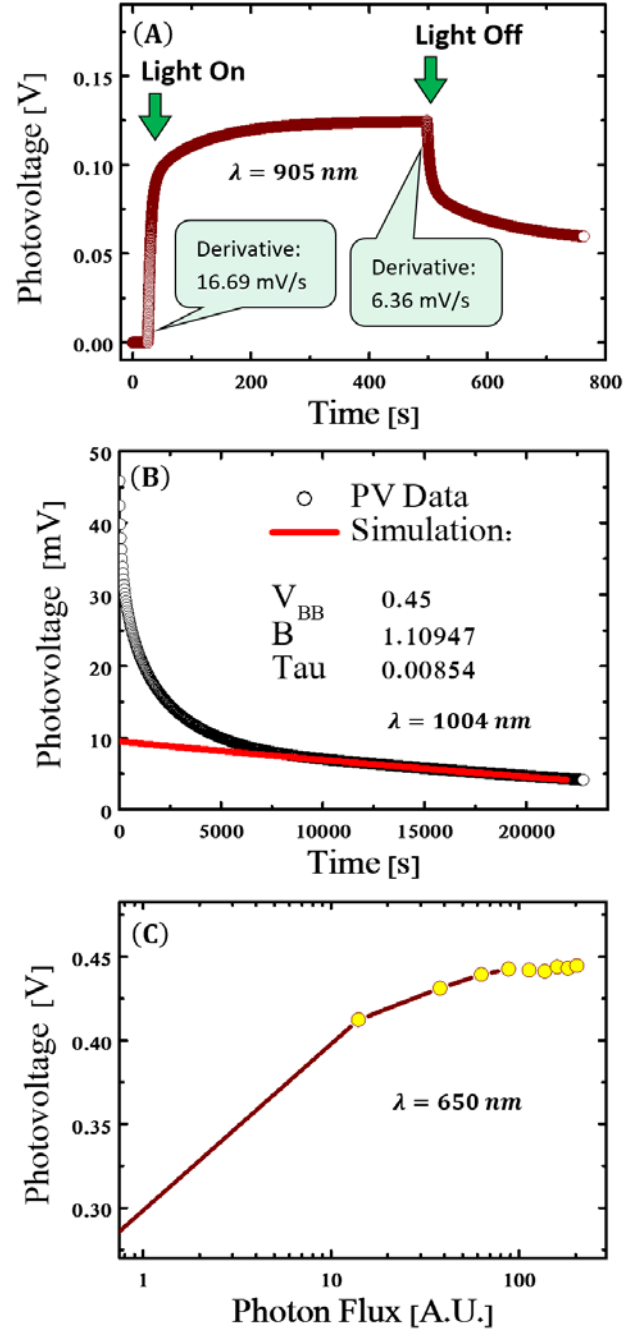

FIG 2 (a) Full photovoltage response (light-on and light-off) was carried out in order to calculate the band banding using derivative methods for comparison. (b) Photovoltage decay after illumination at 1004 nm for 10 s, and simulation. The part where the simulation overlaps the measured data is the range over which our thermionic emission model is valid (the "slow" range). (c) Photovoltage as a function of photon flux was obtained for comparison using a 650-nm laser diode. Saturation is observed slightly below 0.45 V.

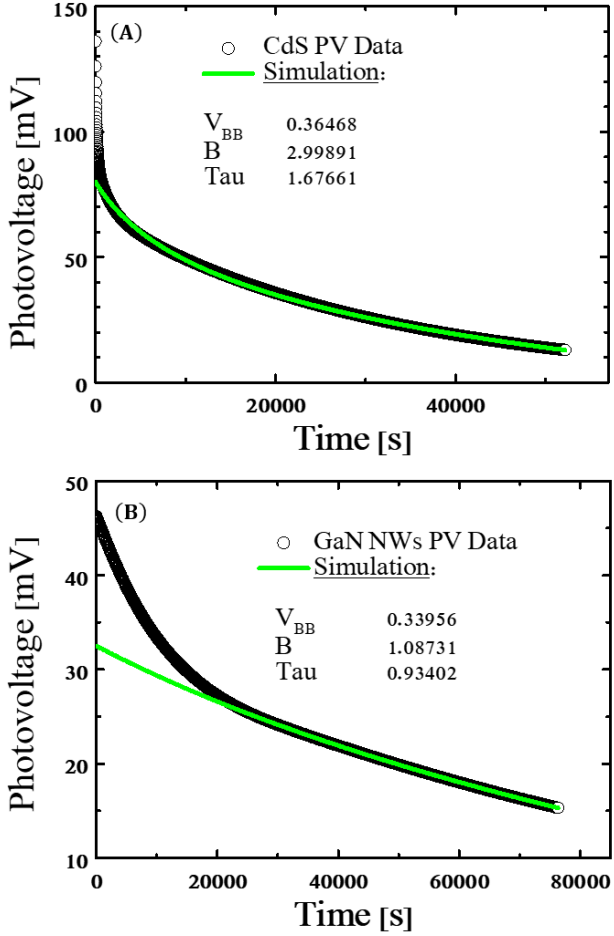

FIG 3 (a) Post-illumination photovoltage decay on the (0001) surface of CdS (data – open circles, simulation - continuous green curve). (b) Post-illumination photovoltage decay on HVPE-grown c-oriented GaN nanowires: data, and simulation.

The GaAs sample was illuminated at 1004 nm. The CdS was illuminated at a wavelength of 535 nm. The GaN wires were illuminated at 375 nm. The measurements were carried out in a dark Faraday cage. Contact potential difference was measured using a vibrating Kelvin probe (Besoke Delta Phi GmbH) on the sample surface. The surface orientations were: GaAs(100) and CdS (0001). The GaN wires grew in the (0001) direction, hence most of the sensed surfaces must have been a-plane surfaces (10-10).

For data fitting, we used the MATLAB software. The response data were fitted using either the Lavenberg-Maequadt, or the Simplex algorithms.

## IV. RESULTS

(1) GaAs wafer

As a thoroughly-studied semiconductor, GaAs seems to make a good test case for the proposed method. Figure 2a a full response (both the light-on and light-off responses). The light-off response is typically orders-of-magnitude longer than the light on response. A typical photovoltage decay acquired from the (100) surface of the GaAs sample after the light was turned off (black open circles) is shown in Fig. 2b. It is comprised of a short and fast drop followed by a long, slow, decay. Another curve (red continuous line) shows a simulated curve calculated using the parameters obtained from the fit. We only attempted to fit the slow part of the curve, because our model does not faithfully describe the fast part. This fit yielded an equilibrium band-bending, $V_{BB} = 0.45 \pm 0.12 V$ (corresponding to surface charge density of $N_{TD} = 1.14 \pm 0.21 \cdot 10^{12} cm^{-2}$), equilibrium ratio of unoccupied to occupied states of $B = 1.11$, and a time constant $\tau = 0.00854\ sec$. Since the Hall effect measured value of the doping, $N_D = 2 \cdot 10^{17}\ cm^{-3}$, we could now calculate the capture cross-section for electrons, $\sigma_n = 6 \cdot 10^{-11} cm^2$.

To compare our results with other methods, the data in Fig 2a was used to calculate the equilibrium band-bending by the methods of Kronik et al. and Reshchikov at al.[10,14]. Using the method of Kronik et al. we obtained $V_{BB} = 77\ mV$, while using the method of Reshchikov et al., we obtained $V_{BB} = 0.1\ mV$. The data of Fig. 2c shows the photovoltage as a function of photon flux (photo-saturation data). The saturation is obtain at a photo-voltage of 444 mV.

(2) CdS wafer

Figure 3a shows photovoltage decay data acquired from CdS (open circles), and the simulated curve based on the parameters from the fit with our model. From this fit, we obtained an equilibrium band bending of $V_{BB} = 0.365\ V$, surface charge density of $N_{TD} = 3.3 \cdot 10^{11}\ cm^{-2}$, and total Surface state density of $B \cdot N_{TD} = 9.9 \cdot 10^{11}\ cm^{-2}$.

(3) GaN Nanowires

Figure 3b shows photovoltage decay data acquired from GaN (open circles), and the fit with our model. From this fit, we obtained an equilibrium band bending of $V_{BB} = 0.34\ V$.

## V. DISCUSSION

In all the above experimental examples, the simulated curves show clearly that the model can only fit the slow part of the photovoltage decay. The deviation observed over the "fast" range, may be a result of several mechanisms, which physics has not been considered in our model.

If we assume that the assumption of Bardeen and Bratain is correct for GaAs, and the bands cannot be bent any further beyond the flat-bend condition, we may use the photo-saturation band-bending, under the reservations of Aphek et al., as a lower bound. Hence, the photo-saturation band-bending may not represent a true flat-band condition, but the actual equilibrium band-bending can only be greater than this value. The values that we obtained using the two derivative-based methods are clearly smaller than this lower bound confirming the frequency band width limitation imposed by the Kelvin-probe in the case of GaAs.

Similar attempt to ours, to explain the surface re-trapping process as thermionic emission has been made by Galbraith and Fisher. [] However, their model had limited validity for several reasons. First, they assumed that the entire response (including the "fast" range) may be described by thermionic emission, while our results suggest otherwise. Second, in their derivation, they assumed a constant junction capacitance, while the capacitance actually varies with the photovoltage. Third, they do not mention the probability of surface trapping, and in practice, this means they actually assumed that it was equal to 1, as in a Schottky barrier, while this probability is not only different than 1 but is also dependent on the photovoltage.

Similar attempts to model the re-trapping process have used various approaches to the problem, but quite unanimously reached the approximated solution of the logarithmic decay with ideality factor correction, probably also because this solution is the only one that appears to fit the entire response curve (including the "fast" range).[7,8,9,10,12,13] However, the fast part of the response may present a rather different physics than the slow part, and therefore attempts to describe the entire curve with a single mechanism may not always be valid.

Indeed, in many cases, additional current, not considered in these models, distorts the slope of the decay curve in a way that makes it impossible to fit with our model. The common practice has been to introduce a "fudge factor" that facilitate the fit. Similar to the common practice in Schottky diode, it has been dubbed "ideality factor". However, unlike the Schottky case, here it may also take values smaller than unity. In fact, what these slope variations mean is that there is an additional mechanism at work which introduces a flow of positive (or negative) charges into the surface states, in addition to the flow of re-trapped electrons described in the model, and this additional flow reduces (or increases) the slope relative to the unity. Since the additional flow is not accounted for by the model, the introduction of the ideality factor only serves the convenience of achieving a better mathematical fit, but the parameters obtained by such fit are in *error*, because the equation no longer describes the assumed physics. Hence, from the physics point of view, the ideality factor actually spoils the fit. As in the case of the Schottky barrier, studies have been carried out in attempt to understand the physics behind the photovoltage decay ideality factor, and correlations of its values with various types of surface state scenarios and various ratios of capture cross-sections have been suggested.[3,27,28]

We mention in passing that in several papers authors have also interpreted the photovoltage decay as a series of exponential decay terms,[29,30] or as a stretched exponential.[31]

Rigorous treatments of the problem have been suggested by Balestra et al.,[6] Kronik et al.,[10] and Reshchikov et al.[14] In all of them the treatment was based on rate equations and the differential equation obtained for the re-trapping process is identical to the one we obtain. Reshchikov et al. provides a solution for the equation for the case of the post-illumination decay. However, in their solution, they still assume a constant, photovoltage-independent, trapping probability, and adopt an approximation, which, in practice, is identical to the constant capacitance assumption of Galbraith and Fisher. While these assumptions carry the clear benefit of reducing the differential equation to an analytical form, they also lead to the widely-accepted solution of logarithmic decay.

In the present treatment, we have taken the hard track, avoiding the above approximations, while taking into account the voltage-dependence of the trapping probability. The differential equations we thus obtained were not analytical, and we had to replace certain parts of the integrands with approximating functions – expansions using exponentials. Nonetheless, these approximations do not seem to compromise the physics or the accuracy of the fit.

An apparent disadvantage of our approach is that by adopting Bethe's thermionic emission model, we also adopted his assumption that the barrier height is constant. This assumption is valid for a Schottky barrier. However, in the presently-studied process, there is no question whether the surface potential barrier does vary in the process of re-trapping. The question is only by how much. Since the variation in the total barrier height is typically much smaller than the change in the band bending, this barrier variation must be negligible over the slow-varying part of the response, where even the change in the band bending is extremely small.

Our purpose in formulating this model was to use it as a basis for a method to characterize electrically the free surface, with the main emphasis on the equilibrium surface band bending. Several methods have been previously proposed to this end. The most extensively used methods to obtain the band bending have been photoelectron spectroscopies [32,33] and the photo-saturation technique. While very useful and reliable, the main drawbacks of photoelectron spectroscopies are that they require ultra-high vacuum, and they suffer from the effect of surface photovoltage.[34] In essence, the illumination used to measure the equilibrium band-bending actually moves the system out of equilibrium, inducing a surface photovoltage. In contrast, the photo-saturation technique is based on the photovoltaic effect and on the prediction of Bardeen and Bratain that sufficiently intense illumination at photon energies above the bandgap may cause flattening of the bands at the surface.[35] However, it has been shown by Afek et al. that the photovoltage may, in some cases, saturate before band flattening is actually achieved. [28]

Two methods for obtaining the equilibrium band bending rely on measurement of photoresponse derivatives.[10,14] These methods are generally based on the assumption that each surface photovoltage data point has been obtained under steady state conditions, or otherwise the measured slopes would not be correct.[1] This assumption may be valid, if the photovoltage is measured using a metal-insulator-semiconductor device, but is not always valid for measurements of free surfaces that are typically carried out using a kelvin probe. Transients of wide gap semiconductors, e.g. GaN or ZnO, are typically slow enough to be followed by a Kelvin probe, but this may not be the case for materials of lower bandgap, such as Si, or GaAs.

Finally, it is also possible to measure the band-bending directly without optical excitation by measuring contact potential difference using a Kelvin probe in the dark. The main source of error in this method is parasitic capacitance. While methods to reduce this error have been thoroughly studied, the only way to absolutely eliminate it is to measure photovoltage.[1]

The method we propose in this paper is a photovoltage-based method that offers the advantage of being suitable for use with a Kelvin probe on free surfaces, because it does not use derivatives, and because it fits only the slow-varying part of the photovoltage decay. However, unlike the photovoltage-derivative-based methods that seem to be independent of the ideality factor, our method works correctly only for "ideal" cases of unity ideality factor.

Our method also assumes a single surface state distribution. In case of more than a single distribution, the difference would typically not be limited to the essence of chemical entity, but would typically be also manifested in dissimilar time-constants. Hence, re-trapping in one state would typically be slower than re-trapping in the other. Therefore, while at the "fast" portion of the response, trapping will take place simultaneously in both states, the "slow" portion of the response would typically consist of trapping in one type alone. This means that in most cases, it may be still possible to obtain the correct equilibrium band-bending by fitting the slow part of the curve.

Limiting the fit to the "slow" portion of the response is, therefore, not only required for the validity of our model, but is also beneficial in several other respects: (1) It is typically slow enough to fit within the frequency bandwidth of the Kelvin probe, (2) it avoids band-to-band excitation and consequential effects, such as the Dember potential, and (3) it typically involves only a single surface state.

The following technical aspects need to be considered when applying the proposed method: (1) If one wishes to obtain the current-voltage characteristics of the free surface, the data should be acquired with high statistics to increase the signal to noise ratio as much as possible, because a derivative will always amplify the noise. (2) The acquired data needs to be a monotonic and continuous function of the measurement time. (3) The acquisition has to be continued until the voltage reaches a steady value. In GaAs, for example, this may be achieved in about an hour. By contrast, ZnO or GaN typically require over 24 hours. (4) The useful data is at a band bending range greater than the thermal voltage, kT/q. This requires that the photon-induced barrier lowering will be away from flat band at least by this value. While this condition is easy to achieve in wide gap materials, it may be challenging, or impossible, in low gap materials. (5) If the photon flux is large, e.g., a powerful laser, at photon energies that exceed the bandgap, part of the band flattening may be due to screening of the field by the excess carriers, in addition to the aforementioned Dember effect.[2] These effects typically decay fast due to band-to-band recombination, and their effect is felt only a short time after the light is turned off (over the non-thermionic range that is not covered by our model). They should be clearly noticed as a deviation from the linear current-voltage curve.

Finally, under certain adjustments, the same approach may be useful in the characterization of temporal responses in other types of junctions.

In summary, we presented a contactless method for electrical characterization of a free semiconductor surface. We used this method to measure the equilibrium surface band bending in layers and nanowires. The method may also yield the equilibrium Fermi level position at the surface, the density of surface states, the density of charge trapped in surface states, the capture cross section for majority carrier traps, and the surface built in field. We also showed that the method may be conveniently used on nanowires avoiding the need to make contacts to individual wires. We claim that this method is not limited to free surfaces but can be used on any junction of a semiconductor with other materials, such as a metal, or an insulator.

## ACKNOWLEDGEMENTS

Financial support from the Office of Naval Research Global through a NICOP Research Grant (No. N62909-18-1-2152) is gratefully acknowledged.

# APPENDIX

## I. Solution of Equation 7 (for bulk layers)

Equation 7 in the manuscript may be solved by separation of variables

$$\frac{exp(Ax)}{(B-\sqrt{x})\sqrt{x}}\frac{dx}{dt} = \frac{1}{\tau} \quad (7)$$

$$\int_{x_0}^{x}\frac{\exp(Ax)}{B\sqrt{x}-x}dx = \frac{1}{\tau}\int_{t_0}^{t}dt \quad (A1)$$

The integral on the left-hand side may be separated into two terms

$$\int_{x_0}^{x}\frac{\exp(Ax)}{B\sqrt{x}-x}dx = \frac{1}{B}\int_{x_0}^{x}\frac{\exp(Ax)}{\sqrt{x}}dx + \frac{1}{B}\int_{x_0}^{x}\frac{\exp(Ax)}{B-\sqrt{x}}dx \quad (A2)$$

The integration of the first term is straightforward

$$\frac{1}{B}\int_{x_0}^{x}\frac{\exp(Ax)}{\sqrt{x}}dx = \quad (A3)$$

$$= \frac{1}{B}\sqrt{\frac{\pi}{A}}\left[erfi(\sqrt{Ax}) - erfi(\sqrt{Ax_0})\right]$$

We note that the imaginary error function, $erfi(x)$, is actually an absolutely real function. We are now left with the second integral, which is not integrable analytically. The integrand is a product of two functions: $f(x) \cdot g(x)$, where

$$f(x) = \frac{1}{B-\sqrt{x}} \qquad g(x) = \frac{\exp(Ax)}{B} \quad (A4)$$

We would like to replace $f(x)$ with another function that will turn the product of the two functions integrable. The function is actually a 2D function of the variables x, and B. x may get values between 0 and 1, while B may be expected to vary between 1 and 3 (the likely values of B are greater than 1, while at low values of x the emission is not thermionic, so we cannot fit there). Figure A1a shows a surface plot of $f(x, B)$. Figure A1b shows the fitting function:

$$f_{fit}(x) = exp\left[\frac{1-B}{2} + \left(0.18 + \frac{22.5}{(5B-2)^2}\right)x\right] \quad (A5)$$

Figure A1c shows the difference between the function and its fit (the fit error).

Now we may integrate $f_{fit}(x) \cdot g(x)$ instead of $f(x) \cdot g(x)$. We get an integrand of the form $exp(a + b \cdot x)$ which is straightforwardly integrable.

The full solution we get is

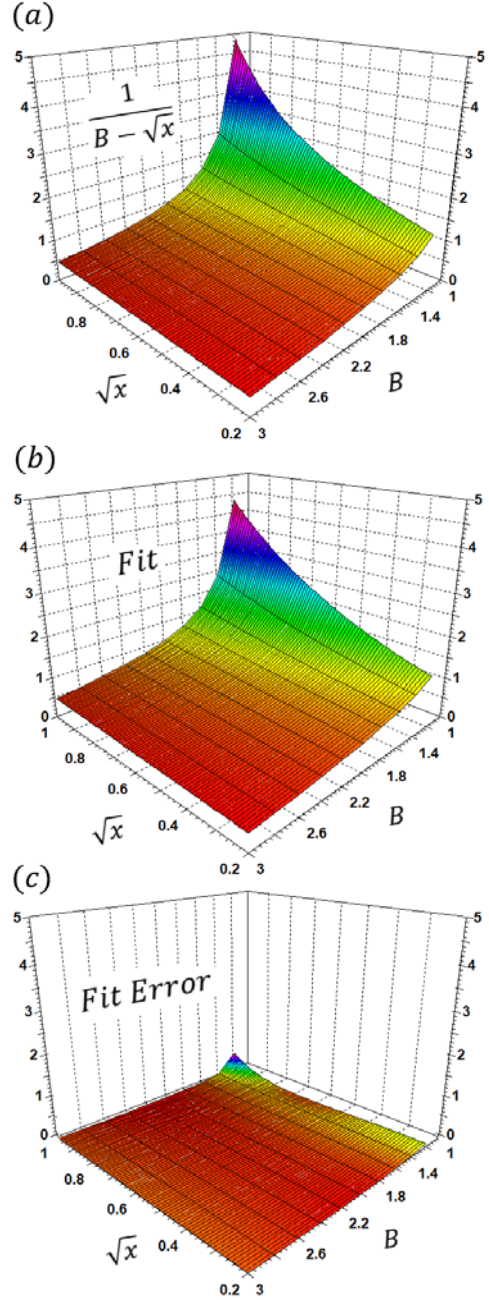

FIG A1 (a) the 2-variable function that we wish to approximate with an exponential function of the same variables. (b) The approximation function. (c) the difference between the original function and its approximation. All curves are shown over the same ranges of x and B values, which are relevant to our problem.

$$\frac{t - t_0}{\tau}$$
$$= \frac{1}{B} \left\{ \frac{exp\left[\frac{1-B}{2} + \left(0.18 + A + \frac{22.5}{(5B-2)^2}\right)x\right]}{\left(0.18 + A + \frac{22.5}{(5B-2)^2}\right)} \right.$$
$$- \frac{exp\left[\frac{1-B}{2} + \left(0.18 + A + \frac{22.5}{(5B-2)^2}\right)x_0\right]}{\left(0.18 + A + \frac{22.5}{(5B-2)^2}\right)}$$
$$\left. + \sqrt{\frac{\pi}{A}} \cdot \left[erfi\left(\sqrt{Ax}\right) - erfi\left(\sqrt{Ax_0}\right)\right] \right\} \quad (A6)$$

**II. Solution of Equation 13 (for nanowires)**

Equation 13 is similar to equation 7 but has only the second integrand

$$\int_{x_0}^{x} \frac{\exp(Ax)}{B - \sqrt{x}} dx = \frac{1}{\tau_2} \int_{t_0}^{t} dt \quad (A7)$$

This problem has already been solved in the previous section:

$$\frac{t - t_0}{\tau_2}$$
$$= \frac{exp\left[\frac{1-B}{2} + \left(0.18 + A + \frac{22.5}{(5B-2)^2}\right)x\right]}{\left(0.18 + A + \frac{22.5}{(5B-2)^2}\right)}$$
$$- \frac{exp\left[\frac{1-B}{2} + \left(0.18 + A + \frac{22.5}{(5B-2)^2}\right)x_0\right]}{\left(0.18 + A + \frac{22.5}{(5B-2)^2}\right)} \quad (A8)$$